\documentclass{aastex}
\usepackage{emulateapj5}
\usepackage{apjfonts}
\input{epsf}
\begin{document}

\newcommand{\lap}{$L_{38}^{-1/3}$}
\newcommand{\ergs}{\rm \su  erg \su s^{-1}}
\newcommand{\etal}{ {\it et al.}}
\newcommand{\porb}{ P_{orb} }
\newcommand{\Po}{$P_{orb} \su$}
\newcommand{\pdot}{$ \dot{P}_{orb} \,$}
\newcommand{\pot}{$ \dot{P}_{orb} / P_{orb} \su $}
\newcommand{\mm}{$ \dot{m}$ }
\newcommand{\mdot}{$ |\dot{m}|_{rad}$ }
\newcommand{\myr}{ \su M_{\odot} \su \rm yr^{-1}}
\newcommand{\msol}{\, M_{\odot}}
\newcommand{\ppp}{ \dot{P}_{-20} }
\newcommand{\cms}{ \rm \, cm^{-2} \, s^{-1} }
\newcommand{\pdott}{ \left( \frac{ \dot{P}/\dot{P}_o}{P_{1.6}^{3}}\right)}
\newcommand{\be}{\begin{equation}}
\newcommand{\ee}{\end{equation}}
\newcommand{\nn}{\mbox{} \nonumber \\ \mbox{} }
\newcommand{\ba}{\begin{eqnarray}}
\newcommand{\ea}{\end{eqnarray}}
\newcommand{\Alfven}{ Alfv\'{e}n }

\def\be{\begin{equation}}
\def\ee{\end{equation}}
\def\p{\phantom{1}}
\def\pmu{\mox{$^{-1}$}}
\def\kms{km^s$^{-1}$}
\def\sbu{mag^arcsec${{-2}$}}
\def\e{\mbox{e}}
\def\dex{\mbox{dex}}
\def\L{\mbox{${\cal L}$}}
\def\gte{\lower 0.5ex\hbox{${}\buildrel>\over\sim{}$}}
\def\lte{\lower 0.5ex\hbox{${}\buildrel<\over\sim{}$}}
\def\loe{\lower 0.6ex\hbox{${}\stackrel{<}{\sim}{}$}}
\def\goe{\lower 0.6ex\hbox{${}\stackrel{>}{\sim}{}$}}

\def\sT{\sigma_{\rm T}}
\def\epsIC{\hbar\omega_{\rm IC}}
\def\epsX{\epsilon_X}

\shorttitle{High-Energy Emission from Magnetars}
\shortauthors{Thompson and Beloborodov}
\title{High-Energy Emission from Magnetars}

\author{
C. Thompson\altaffilmark{1} and
A. M. Beloborodov\altaffilmark{2,3}
}
\altaffiltext{1}{Canadian Institute for Theoretical Astrophysics,
60 St. George St., Toronto, ON M5S 3H8}
\altaffiltext{2}{Department of Physics, Columbia University}
\altaffiltext{3}{Astro-Space Center of Lebedev Physical Institute,
Profsojuznaja 84/32, Moscow 117810, Russia}
\submitted{{\it Astrophysical Journal}, 634, 565, 2005}

\begin{abstract}
The recently discovered soft gamma-ray emission from the anomalous
X-ray pulsar 1E~1841-045 has a luminosity 
$L_\gamma\sim 10^{36}$~ergs~s$^{-1}$.  This luminosity exceeds the 
spindown power by three orders of magnitude and must be fed
by an alternative source of energy such as
an ultrastrong magnetic field.  A gradual release of energy in
the stellar magnetosphere is expected if it is twisted and
a strong electric current is induced on the closed field lines.
We examine two mechanisms of $\gamma$-ray emission associated with 
the gradual dissipation of this current.
(1) A thin surface layer of the star is heated by the downward 
beam of current-carrying charges, which excite Langmuir turbulence
in the layer.  As a result, it can reach a temperature 
$k_BT \sim 100$~keV and emit bremsstrahlung photons up to this
characteristic energy.
(2) The magnetosphere is also a 
source of soft $\gamma$-rays at a distance of $\sim 100$ km from
the star, where the electron cyclotron energy is in the kilo-electron
volt range.
A large electric field develops in this region
in response to the outward drag force felt by the current-carrying 
electrons from the flux of kilo-electron volt photons leaving the star.
A seed positron injected in this region undergoes a runaway
acceleration and upscatters X-ray photons above the threshold for pair
creation. The created pairs emit a synchrotron 
spectrum consistent with the observed 20-100~keV emission. This 
spectrum is predicted to extend to higher energies and reach a 
peak at $\sim 1$ MeV.

\end{abstract}


\section{Introduction}\label{secone}

The Soft Gamma Repeaters and Anomalous X-ray Pulsars share several
characteristics:  persistent X-ray luminosities in the range 
$10^{34}-10^{36}$ erg/s; spin periods between 5 and 12 s;
characteristic ages $P/\dot P \sim 10^3-10^5$ yrs; 
the release of short ($\sim 0.1$ s) and spectrally hard X-ray
bursts; and an  inferred magnetic field of $10^{14}$-$10^{15}$~G.
(See Woods \& Thompson 2004 for a review.)
The persistent X-ray emission of these ``magnetars''
has, until recently, been measured in the 1-10 keV band.  
Within that band,
it can typically be modeled by a superposition of a 
thermal (black body) component ($k_BT_{\rm bb} \sim 0.4$ keV)
and a power law with a photon index in the range 
$-4$ to $-2$.
 The hardness of the power law correlates with the overall activity
 in outbursts, and with the spindown torque (Marsden \& White 2001). 

Only recently has there been a convincing measurement of the
persistent emission of a magnetar above 20 keV 
(Kuiper, Hermsen, \& Mendez 2004).  RXTE and Integral
revealed a hard, rising energy spectrum between
$\sim 20$ and 100 keV in the  AXP 1E 1841$-$045, which dominates
the apparent bolometric output.  We focus in
this paper on the important implications of this detection for the mechanism
that supplies the persistent non-thermal emission of the AXPs and SGRs.

The activity of a magnetar 
is ultimately powered by the decay of an ultrastrong magnetic field
(Thompson \& Duncan 1996).  
Several properties of their persistent emission 
-- the  non-thermal spectrum, and the long-lived changes 
in $\dot{P}$ and pulse profile following outbursts 
(Woods et al. 2001; Kouveliotou et al. 1998; Woods et al. 2002)
-- are all consistent with a non-potential (twisted) magnetosphere
(Thompson, Lyutikov, \& Kulkarni 2002, hereafter TLK).
A twisted internal magnetic field 
provides a plausible (repeating) source
of magnetic helicity for the exterior of the star
by shearing the crust.

Magnetars are slowly rotating and their light cylinders have large radii,
$R_{\rm lc} \equiv c/\Omega \sim 3\times 10^{10}$ cm (about 30,000
times the neutron star radius, $R_{\rm NS} \sim 10^6$ cm).   As in radio
pulsars, the rotation of the star
induces a net magnetospheric charge density
$\rho_{\rm co} = -{\bf\Omega}\cdot{\bf B}/2\pi c$ (Goldreich
\& Julian 1969).  This drives a current flowing on the open magnetic
lines, which  extend from a small part of the stellar surface out
to the distant light cylinder. 
The currents induced by a twisting motion of the crust
are much stronger, and may flow everywhere inside the magnetosphere.  

We focus on the region well inside the light cylinder, and make
use of a simple model (TLK) to investigate the dissipative properties 
of such currents.   
The strength of the magnetic field is characterized by the surface 
polar field $B_{\rm pole}$.  
The shape of the field lines depends on 
a single parameter:  the angle $\Delta\phi_{\rm N-S}$ through which
the field lines anchored close to the magnetic poles are twisted.
The poloidal field remains close to a pure dipole
as long as $\Delta\phi_{\rm N-S}\la 1$ radian.  Outside the star,
the current density has the form 
\be\label{curden}
{\bf J} = {c\over 4\pi}{\bf\nabla}\times{\bf B} \simeq
{c {\bf B}\over 4\pi r}\sin^2\theta\,\Delta\phi_{\rm N-S}\;\;\;\;\;\;
(r > R_{\rm NS}).
\ee 

The current must be driven by an induced electric field against the force
of gravity; a dynamic model of the current is developed in an upcoming paper 
(A. M. Beloborodov \& C. Thompson, in preparation). 
Charges of opposite signs are 
lifted at the two footpoints of a closed magnetic line, travel along the line,
and return to the star at the opposite end.  Far inside the light cylinder,
the co-rotation charge density  is much smaller than the absolute 
charge density $J/\beta_\parallel c$ supplied by the current,
\be\label{GJcomp}
{\rho_{\rm co}\over J/\beta_\parallel c} \sim \beta_\parallel\,
\left({R_{\rm max}\over R_{\rm lc}}\right).
\ee
Here, $\beta_\parallel c$ is a characteristic particle drift speed 
along the field lines,
and $R_{\rm max}(r,\theta) = r/\sin^2\theta$ is the maximum extent of
a field line (when the poloidal field is nearly dipolar, $\Delta\phi_{\rm N-S}
\la 1$).  
In contrast with the open field lines, the plasma is nearly neutral, 
and each end of the circuit receives a flow of charges from the opposite
end.  In this situation a small polarization of the magnetospheric plasma can 
maintain the modest co-rotation charge density. 
Therefore, the formation of charge-starving 
gaps with strong electric fields is not necessary to maintain the
closed-field current.

The stability of such a large-scale, non-potential magnetic field
is closely related to Taylor's conjecture for the relaxation of dilute
plasmas containing a helical magnetic field (Taylor 1986).
A twisting up of the external magnetic field involves a transfer
of magnetic helicity from the interior of the star to its exterior.
Although small-scale irregularities will be damped rapidly by reconnection
(and can manifest themselves as SGR bursts), several lines of observational
evidence indicate that the global helicity 
decays on a much longer resistive timescale.  
Persistent non-thermal X-ray emission is observed from
a number of AXPs and SGRs over a period of a decade or longer, accompanied
by little X-ray burst activity.  There is 
no evidence for an undetected population of low-energy
seismic events which would occur independently of SGR burst activity:
the cumulative energy radiated in SGR bursts diverges on the high-energy 
tail of the burst energy distribution (e.g. G{\" o}{\u g}{\" u}{\c s} 
et al. 1999). In such a situation, only
toroidal field energy that is dissipated {\it outside} the star will
be converted efficiently to energetic particles and non-thermal X-rays; 
field energy dissipated deep in the crust will be converted largely to heat.
Dissipation (resistive)
timescales as long as $\sim 10$ years are implied in some AXPs
by the observation of persistent non-thermal X-ray emission without
any concurrent burst activity.

\section{Electron Acceleration, Pair Creation and Surface Heating}\label{paircas}

A small electric field must be present to drive the current against
the force of gravity.  
If the positive charges lifted at the anode end of the field line 
are protons (mass $m_p$), the characteristic energy\footnote{In
comparison,  the electric potentials generated by
a fully charge-separated plasma of particle density $\sim J/ec$ would
be much stronger, by the ratio $\sim \Delta\phi_{\rm N-S}eBR_{\rm NS}^2/GMm 
= 10^{15}\,\Delta\phi_{\rm N-S}\,B_{15}(m/m_p)$.  
Note that, in this paper, we use
the normalization $X = X_n \times 10^n$ for quantity $X$, as measured 
in c.g.s. units.}   per charge in
such a ``minimal-resistance'' circuit is
$e|\Delta\Phi| \sim (GMm_p/R_{\rm NS})(1-R_{\rm NS}/R_{\rm max})
= 200\,(1-R_{\rm NS}/R_{\rm max})$ MeV.  
 This energy is dissipated by plasma instabilities along the circuit,
in particular when the charges return to the star and hit its surface 
layer.
Integrating the current density $J$ (eq. [\ref{curden}]) over the 
surface of the star,  one obtains the net current $I$ and the 
corresponding dissipation rate 
$L_{\rm diss}\sim (I/e)(GMm_p/R_{\rm NS})$.  This is
$\sim 10^{35}-10^{36}$~ergs s$^{-1}$ for a moderately strong twist,
comparable to but not exceeding the observed X-ray luminosities
of SGRs and AXPs.   Potential drops $e|\Delta\Phi| \gg 200$~MeV
would be inconsistent with observed luminosities and imply
a short dissipation timescale, excluding long-living twists.
If ions carry a significant fraction of the positive charge in the
circuit, one concludes that electrostatic potentials must be
present which could accelerate electrons to high Lorentz factors,
\be
\gamma_e \sim \gamma_e({\rm e-i}) \equiv 
{GM_{\rm NS} m_p\over R_{\rm NS} m_ec^2} \simeq 400.
\ee

Alternatively, the upper layers of the neutron star atmosphere
could be hot enough that electron-positron pairs contribute significantly
to the pressure, and therefore to the current flow above the surface of 
the star.  In this case, most of the dissipation 
along the circuit must be concentrated within the atmosphere:
the gravitational binding energy of an electron or positron to the 
star is only $\sim 100$ keV.   The observed high energy emission would
then be a direct consequence of the mechanism that supplies charges
to the magnetosphere.  Indeed, the number flux of charged particles
that carry the current is only $\sim 10^{-2}
\varepsilon_J B_{15}$ of the flux of 100 keV photons (if these
photons originate at the surface of the star).  Here,
we have expressed the current density as
$J = \varepsilon_J (cB_{\rm NS}/4\pi R_{\rm NS})$, where
$\epsilon_J = \Delta\phi_{\rm N-S}\sin^2\theta$ from eq. (\ref{curden}).

A fast magnetospheric flow of charges must deposit a significant fraction
of its energy in a thin, upper layer surface layer of the atmosphere. 
(We assume that the surface is non-degenerate and
composed of light elements (H or He) in what follows.)
The downward electron beam 
at the anode end of the field line\footnote{A downward positron beam
at the cathode surface would behave similarly.} will be subject to a strong
beam instability.  The initial growth length of this instability is 
small compared with the vertical hydrostatic scale of the target 
plasma, being a multiple of the plasma scale $c/\omega_{Pe}$, where
$\omega_{Pe} = (4\pi n_e e^2/m_e)^{1/2}$ is the electron plasma frequency.
For example, when the beam has a high Lorentz factor $\gamma_e$,
the growth length is 
$\ell_{\rm beam} \sim \gamma_e\,(n_{\rm beam}/n_e)^{-1/3}\,c/\omega_{Pe}$.
Flattening of the electron distribution
function at large (relativistic) momentum implies the deposition of
a minimum of ${1\over 2}$ of the beam energy into plasma oscillations. 

The current will also be subject to anomalous resistivity in
the uppermost layer of the atmosphere.  Rapid cyclotron cooling
keeps the temperature of the ions well below the surface cyclotron
energy of $\hbar\omega_{c,p} = \hbar eB/m_pc = 6.3\,B_{15}$ keV
(for protons).  The ion temperature is therefore typically much
smaller than the electron temperature at the base of the beam-heated layer,
and strong ion-sound turbulence may be excited.  The resistive
heating is concentrated in the upper layers of the atmosphere, 
where the electron drift speed $|J|/en_e$ exceeds the ion sound speed
$(T_e/m_p)^{1/2}$.  Here the electrons are effectively collisionless.

We now determine the electron temperature deeper in the atmosphere,
where the electrons are collisional.  This corresponds to electron
columns larger than  $N_{\rm col} \sim 
\sigma_{\rm Coulomb}^{-1} \sim \sigma_T^{-1} (kT_e/m_ec^2)^2$,
where $\sigma_{\rm Coulomb}$ is the Coulomb cross section.
Because the luminosity of the beam is far below the Eddington luminosity,
most of the deposited energy will be conducted downward by the electrons
to a Thomson depth $\tau_T \sim 1$, where cooling is effective.
The downward conductive heat flux is\footnote{The Coulomb logarithm 
is about unity when all the electrons are confined to their 
lowest Landau state.}
\be\label{hcond}
Q^-_{\rm cond} \simeq -{ck_B\over\sigma_T}\,
\left({k_BT_e\over m_ec^2}\right)^{5/2}
{dT_e\over dr} < 0.
\ee
We assume that the heated plasma is hydrostatically supported 
by electron pressure, which gives $k_BdT_e/dr\sim gm_p$ and
$Q^-_{\rm cond} \simeq -(m_pc g/\sigma_T)\,
(k_BT_e/m_ec^2)^{5/2}$.
We balance $Q^-_{\rm cond}$ with the rate of surface heating,
\be\label{hheat}
Q^+_{\rm heat} = {1\over 2}{J\over e}(\gamma_e m_ec^2),
\ee
where $\gamma_e m_ec^2 = e|\Delta\Phi|$ is the characteristic
energy dissipated per current-carrying charge.
Setting $Q^-_{\rm cond} = -Q^+_{\rm heat}$
gives the characteristic conduction temperature
\ba\label{tmin}
  k_B T_Q
&\simeq&
\left({\varepsilon_J\over 3}\alpha_{\rm em} 
              \frac{B_{\rm NS}}{B_{\rm QED}}\right)^{2/5}\,
\left[{\gamma_e\over \gamma_e({\rm e-i})}\right]^{2/5}\;m_ec^2\nn
&\sim& 160\;(\varepsilon_JB_{\rm NS,15})^{2/5}\,
\left[{\gamma_e\over \gamma_e({\rm e-i})}\right]^{2/5}\;\; {\rm keV}.
\ea
Here $\alpha_{\rm em} = 1/137$.  

Photons emerging from deep in the star cannot cool the surface 
layer because they are in the extraordinary polarization mode ($E$-mode,
see, e.g., Sil'antev \& Iakovlev 1980). The scattering of this mode 
is suppressed by a factor $\sim (\omega m_ec/eB)^2$ 
and the mode couples weakly to the beam-heated layer.  (When $B > 10^{14}$ G,
the interconversion of E-mode to O-mode photons only occurs deep
in the star's atmosphere, where the O-mode photons are effectively
trapped:  Lai \& Ho 2003.)
The layer must then cool by its own emission, and 
most of the energy deposited
at the surface will be radiated by ordinary mode (O-mode) photons.  
The scattering and absorption opacities of this mode
are comparable to those in an unmagnetized plasma.  For example,
the scattering depth encountered by an O-mode photon is
$\tau_O \simeq \tau_T\sin^2\theta_{kB}$, which is $\tau_O \simeq \tau_T$
after averaging over angles.

We have calculated the free-free emission from this beam-heated
atmosphere at depths greater than $\tau_T \sim \sigma_T N_{\rm col}
\sim 10^{-1}$, given the downward heat flux estimated above.  The
free-free emissivity
$\varepsilon_{\rm ff} \simeq (2/\pi)^{3/2}\,\alpha_{\rm em}\,
(k_BT_e/m_ec^2)^{1/2}\,n_e^2\sigma_T m_ec^3$ is small at the top
of the collisional portion of atmosphere, as
is the Compton parameter of the O-mode photons,
$y_O = 4\,{\rm max}(\tau_O,\tau_O^2)\,(k_BT_e/m_ec^2)$.
Deeper in the atmosphere, one can estimate the temperature at which
the electrons cool down by setting
$\varepsilon_{\rm ff} \Delta r = Q^-_{\rm cond}$, where
$\Delta r \simeq k_BT_e/m_p g \sim 10^3$ cm is the vertical scale height.
One finds 
\be\label{tcool}
k_BT_{\rm brems} \;=\; (2\alpha_{\rm em}\tau_T^2)^{1/3}m_ec^2
\;=\; 120\,\tau_T^{2/3}\;\;\;{\rm keV}.
\ee

The O-mode emission from this hydrostatic, conducting layer has
been summed numerically.  The spectrum is well-approximated
by thermal bremsstrahlung with a temperature 
$k_BT_{\rm brems} = 80$ keV for $\varepsilon_J B_{15}\,
[\gamma_e/\gamma_e({\rm e-i})] = 2$ \{scaling as
$k_BT_{\rm brems} \sim 
[\varepsilon_J B_{15}\,\gamma_e/\gamma_e({\rm e-i})]^{2/5}$\}.  As shown
in Fig. 1, the photon index is
very hard ($\simeq -1$) below an energy $\sim k_BT_{\rm brems}$.  
The calculated spectrum has a slightly more extended high energy tail
than the pure bremsstrahlung spectrum.

We have generalized this calculation to include the Compton heating
of the radiation as it passes through upper, hotter layers, and
find a nearly identical result.  The equation governing the conductive
heat flux now becomes
\be\label{condevol}
{dQ^-_{\rm cond}\over dr} = -\varepsilon_{\rm ff} - 
     {4k_{\rm B}[T_e(r)-\langle T_\gamma \rangle]\over m_ec^2}
          \sigma_T n_e c U_\gamma,
\ee
where $U_\gamma \simeq (1+3\tau_T)(-Q^-_{\rm cond})/c$ is the energy density
in the bremsstrahlung radiation at Thomson depth $\tau_T$.
A bremsstrahlung source
spectrum of temperature $T_{\rm brems}$ is modified by Compton scattering
according to $(T_{\rm brems})^{-1}\,dT_{\rm brems}/dt
= 4(n_e\sigma_T c)k_{\rm B}(T_e-T_{\rm brems})/m_ec^2$.
The emergent X-ray spectrum is obtained by summing the contributing
from each emitting layer, after including the effects of Compton
scattering by the upper, hotter layers.  The downward heat flux
is diminished by Compton cooling, the effects of which are incorporated into
eq. (\ref{condevol}) using a mean temperature  $\langle T_\gamma\rangle 
= [-d\ln(U_\gamma)/dE_\gamma]^{-1}$ near the spectral peak.  This
mean temperature evolves according to 
$\langle T_\gamma\rangle^{-1} d\langle T_\gamma\rangle/d\tau_T
= -4(1+3\tau_T)k_{\rm b}(T_e-\langle T_\gamma\rangle)/m_ec^2 +
\varepsilon_{ff}(T_e-\langle T_\gamma\rangle)/T_en_e\sigma_T Q^-_{\rm cond}$.

\vskip .1in
\centerline{{
\vbox{\epsfxsize=8.4cm\epsfbox{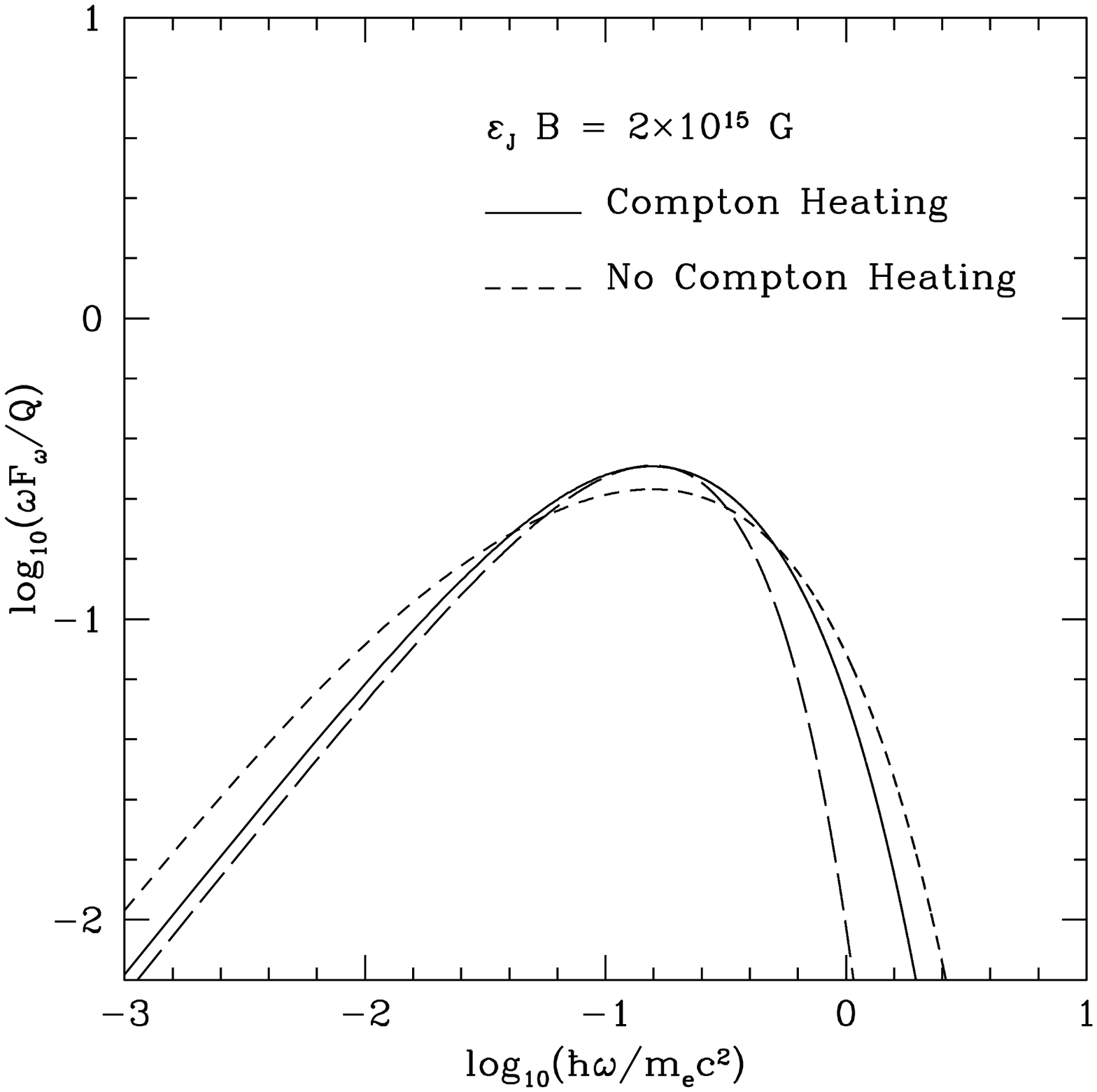}}
}}
\figcaption{X-ray spectrum emerging from a thermal, hydrostatic,
light-element atmosphere which is heated at its surface at a rate
$Q$ per unit area.   The current is normalized
to $\varepsilon_J B_{15} = 2$.  A pure bremsstrahlung emission
spectrum at $T_e = 80$ keV is shown for comparison (long-dashed curve).
}
\vskip .3in

It should be emphasized that hard O-mode photons will be
able to penetrate the strong magnetic field without splitting.
In this way, it is possible to hide most of the energy of
the magnetospheric beam in emission above $\sim 20$ keV
(such as is observed in 1E 1841$-$045; Kuiper et al. 2004). 

Our model differs from the
classic calculations of the X-ray emission of accreting, non-magnetic neutron
star by Zel'dovich \& Shakura (1969) and
Shapiro \& Salpeter (1975) in that i) heat is deposited directly into
the electrons in a thin surface layer, and transported downward mainly 
by electron conduction (which is faster than advection by a factor
$\sim R_{\rm NS}/\Delta r$); and 
ii) the heated atmosphere is not cooled by the blackbody radiation of the 
star; the cooling (O-mode) photons are produced in the atmosphere itself 
at a low free-free optical depth.

The surface heating in our model results from the acceleration of a 
relativistic electron beam in the magnetosphere, combined with strong
and localized anomalous resistivity.  The acceleration is 
governed by electric fields which in turn are determined by the charge 
density distribution. The self-consistent problem of plasma dynamics in 
the electric circuit is addressed in a separate paper (A.M. Beloborodov \& 
C. Thompson, in preparation). Creation of $e^\pm$ pairs plays a crucial role
in this dynamics. The presence of positrons is also important for the 
radiative mechanism investigated in the following section, and here we 
discuss briefly $e^\pm$ production.

First, if the temperature $T_e$ in the upper atmospheric layers reaches 
$\sim 300$~keV/$k_{\rm B}$, then $e^+$-$e^-$ pairs are created directly 
in the atmosphere by the conversion of thermal bremsstrahlung photons off 
the magnetic field. Because these photons are created in the ordinary mode, 
they have a threshold energy for pair creation of 
$\sim 2\,m_ec^2/\sin\theta_{kB}$, where $\theta_{kB}$ is the angle of 
propagation with respect to the magnetic field.  Effective pair 
creation at high $T_e$ tends to regulate the upper cutoff of the persistent 
X-ray spectrum at $\sim 300$ keV, with a weak dependence on the strength 
of the current.

Second, a large output $L_O \sim 10^{36}$~ergs~s$^{-1}$ in $> 100$~keV 
photons from the surface of the star leads to pair creation in the 
magnetosphere. The X-rays are upscattered by relativistic electrons at 
their first Landau resonance and become gamma-rays. (In strong magnetic 
fields, curvature emission of gamma-rays requires much higher energies, 
$\gamma_e \sim 10^7$, than does resonant upscattering; e.g. Hibschman \& 
Arons 2001.) The resonance condition is 
$\hbar\omega_X'\simeq (B/B_{\rm QED})m_ec^2$ for a photon of
frequency $\omega_X'$ that moves nearly parallel to the magnetic field in
the rest frame of the scattering charge. (The resonance condition remains
the same in super-QED magnetic fields due to the effect of the electron
recoil.)  Thus, the electron Lorentz factor required to resonantly upscatter
an X-ray photon of frequency $\omega_X$ is
\be
   \label{gamres} \gamma_e({\rm res})  \simeq 10^2\,B_{\rm 15}\,
           \left({\hbar\omega_X\over{\rm 100 keV}}\right)^{-1}.
\ee
When the magnetic field is $\sim 10\,B_{\rm QED}$ or stronger, the photon
emitted by the excited electron will either directly convert to an
electron-positron pair (if it is emitted in the O-mode), or create a pair
after undergoing the QED process of splitting (if emitted in the E-mode).

\section{Effects of Cyclotron Scattering at $\sim 100$ km}
\label{cycpot}

The 2-10 keV X-ray output of the SGRs and AXPs is high enough
that current-carrying electrons feel a strong drag force through
their cyclotron resonance.  At a distance from the star ($50-200$ km)
where the resonance is in the keV range, this radiative force is
much stronger than the force of gravity (TLK).
In the presence of a persistent electric current, negative charges
(electrons) must return to the star in one magnetic hemisphere.
The force parallel to the magnetic field must therefore be compensated
by an electric force,
\be\label{eq:bal}
F_{\rm rad\,\parallel} = e E_\parallel.
\ee

To calculate the radiative force, we approximate
the X-ray photon field as being radial at $\sim 100$ km.  
The angle $\theta_{kB}$ between an electron and incident photon is
$\cos\theta_{kB} = \hat B\cdot\hat r \simeq 
2\cos\theta/\sqrt{1+3\cos^2\theta}$.
Here $\theta$ is the magnetic polar angle and we have assumed a
nearly dipole geometry  (a weak twist).  An electron at rest
absorbs (unpolarized) X-ray photons with a cross section
$\sigma_{\rm res} = (\pi^2e^2/m_ec)\,\delta(\omega-\omega_{c,e})
(1+\cos^2\theta_{kB})$.
Integrating through the cyclotron resonance $\omega_{c,e} = eB/m_ec$,
the optical depth is $\tau_{\rm res} \sim 
\alpha_{\rm em}^{-1}(\hbar\omega_{c,e}/m_ec^2)^{-1}\sigma_T n_e r$.

The drag force is minimal when the electrons advance slowly to the 
star with a speed $\beta_\parallel \ll 1$.  (They then supply an
optical depth $\tau_{\rm res} \sim \Delta\phi_{\rm N-S}/\beta_\parallel$.)
Integrating over frequency gives
\be\label{frad}
F_{\rm rad\,\parallel} \simeq {\pi^2e\over B}
\left({\omega L_\omega\over 4\pi r^2c}\right)_{\omega_{c,e}}
\cos\theta_{kb}(1+\cos^2\theta_{kB}),
\ee
directed outwards.
The potential accumulated along a given magnetic field line is given by
$e\Delta\Phi = -\int_0^{\pi/2} e E_\parallel (dl/d\theta)\,d\theta$.
It depends on the spectral distribution of the X-ray continuum 
emerging from closer to the star.  For simplicity, we take
a flat energy spectrum, $\omega L_\omega =$ constant, and
normalize the frequency to the cyclotron frequency $\omega_{c,e}(R_{\rm max})$
at the maximum radius $R_{\rm max}$ of the field line in question,
$R_{\rm max}/R_{\rm NS} = 
[eB_{\rm pole}/2m_e c \omega_{c,e}(R_{\rm max})]^{1/3}$.
One finds,
\be\label{potent}
{e|\Delta\Phi|\over m_pc^2} 
= 0.3\,{(\omega L_\omega)_{35}\over B_{\rm pole,15}^{1/3}\,R_{\rm NS,6}}\,
  \left[{\hbar\omega_{c,e}(R_{\rm max})\over {\rm keV}}\right]^{-2/3}.
\ee
This is, in order of magnitude,
$e|\Delta\Phi|\sim\sigma_{\rm res} (\omega L_\omega) /4\pi c r
\sim \tau_{res}(\omega L_\omega)/4\pi n_e r^2 c$.
Notice that the electrostatic potential is proportional to the
X-ray luminosity, and depends weakly on $B_{\rm pole}$.

\subsection{Runaway Positrons and Synchrotron Emission}
\label{pairun}

A strong electrostatic
potential, with a depth of hundreds of MeV,
develops at a radius of $\sim 100$ km in the current-carrying magnetosphere
of a magnetar.  This has dramatic and observable consequences for the
gamma-ray emission  -- if at least a modest fraction of the outgoing
positive charges are positrons.  

In the absence of radiation drag, positrons would quickly become
relativistic in the electrostatic potential (\ref{potent}).
Drag at the cyclotron resonance is ineffective if the resonance sits
at or below the $\sim 1$ keV blackbody peak of the surface X-ray
emission.  To see this, note that the number density of resonant photons,
of energy $\hbar\omega_{\rm res} \sim (1~{\rm keV})\gamma_{e^+}^{-1}$,
scales as $n_\gamma(\omega_{\rm res})
\sim \omega_{\rm res}^2 \sim \gamma_{e^+}^{-2}$. 
The drag force therefore decreases as 
$\gamma_{e^+}^2 (\hbar\omega_{\rm res})n_\gamma(\omega_{\rm res}) 
\propto \gamma_{e^+}^{-1}$,
and resonant scattering of the low-energy photons does not prevent
a runaway acceleration.

The accelerated positrons will not, however, gain the full
electrostatic potential because
their momentum is limited by
{\it non-resonant} inverse-Compton scattering.  The optical depth
for a positron to scatter a thermal X-ray photon is
\be
{\sigma_T L_\omega\over 4\pi r_{\rm res} \hbar c}
= 6.2\,\left({\omega L_\omega\over 10^{35}~{\rm ergs~s^{-1}}}\right)\,
\left({\hbar\omega\over {\rm keV}}\right)^{-2/3}\,B_{\rm pole,15}^{-1/3},
\ee
at the equatorial radius $r_{\rm res}$ where $\omega_{c,e} = \omega$.
Thus, an accelerated positron will scatter keV photons with a
mean free path $\lambda/r_{\rm res} = 1/\sigma_T n_{\rm keV} r_{\rm res}
\sim 0.16\,(\omega L_\omega)_{35}^{-1}\,B_{\rm pole,15}^{1/3}$
(where $n_{\rm keV}$ is the number density of the target X-ray photons).

The energy of the upscattered photons $\epsIC$ can be estimated from a
simple argument. The balance between Compton drag and electrostatic
acceleration reads $\epsIC\simeq eE_\parallel\lambda$.
On the other hand, eq. (\ref{frad}) implies that
$eE_\parallel =\sigma_{\rm res} n_{\rm keV} \hbar\omega$
with $\sigma_{\rm res} \sim \pi^2e^2/m_e c \omega$. 
This gives
\be
\epsIC\simeq \left({\sigma_{\rm res}\over \sigma_T}\right)\hbar\omega
= \frac{3\pi}{8\alpha_{\rm em}} m_ec^2 \sim  100 \;\; {\rm MeV}.
\ee

A more accurate estimate takes into account the spectrum and 
angular distribution of the target X-ray photons.  
The drag force on the positrons
is mainly due to an isotropized component of the X-ray flux,
when the current-carrying electrons supply an optical depth 
$\tau_{\rm res} \sim 1$.   We assume that the input X-ray spectrum has
a Rayleigh-Jeans cutoff at frequencies lower than
$\omega_{\rm bb} \sim 1$ keV/$\hbar$.  In parts of the magnetosphere
where the resonant
frequency $\omega_{c,e}$ sits below $\omega_{\rm bb}$, the positron
reaches the equilibrium energy
$\langle \gamma_{e^+}\rangle^2 \simeq (\varepsilon_{c,e}/\alpha_{\rm em}
\tau_{\rm res})\,(\hbar\omega_{c,e}/m_ec^2)^{-1}$.  This is independent
of the normalization of the X-ray flux,
$\langle\gamma_{e^+}\rangle  \sim 290\,
(\varepsilon_{c,e}/\tau_{\rm res})^{1/2}
(\hbar\omega_{\rm c,e}/{\rm keV})^{-1/2}$.
The factor $\varepsilon_{c,e}$ is the ratio
of $\omega L_\omega$ at the frequency $\omega_{c,e}$, to the luminosity
integrated up to a frequency $\omega_{\rm KN}$ where non-resonant
scattering is Klein-Nishina suppressed.  
One generally finds that $\omega_{\rm KN}$ equilibrates to a value
near $\omega_{\rm bb}$ when $\tau_{\rm res} \sim 1$.  In this case
$\varepsilon_{c,e} \simeq  3\,(\omega_{c,e}/\omega_{\rm bb})^3$.
The equilibrium energy of the positron is independent of the normalization
of the X-ray flux,
$\langle\gamma_{e^+}\rangle \simeq 290\,
 (\varepsilon_{c,e}/\tau_{\rm res})^{1/2}\,
({\hbar\omega_{c,e}/{\rm keV}})^{-1/2}$.  
The inverse-Compton photon has an energy
$
\hbar\omega_{\rm IC} \simeq {4\over 3}\gamma_{e^+}^2\hbar\omega 
\sim 300\,\tau_{\rm res}^{-1}(\omega_{c,e}/\omega_{\rm bb})^2
$ MeV.

The threshold energy for creating $e^+$-$e^-$ pairs directly off the 
magnetic field is $\hbar\omega_{\rm IC}\sin\theta_{kB} 
\sim 0.06\,(\hbar\omega_{c,e}/
m_ec^2)^{-1}m_ec^2 \sim 15$ MeV.  This threshold is exceeded only
in a restricted part of the magnetosphere: the accelerating potential 
is too weak where $\omega_{c,e} \la 0.5\omega_{\rm bb}$.
The resulting
prompt synchrotron radiation peaks at 
$\hbar\omega_{\rm synch}
\sim 5\times 10^{-3}\hbar\omega_{\rm IC}$. This is,
\be
\hbar\omega_{\rm synch} \sim 
1.5\,\tau_{\rm res}^{-1}(\omega_{c,e}/\omega_{\rm bb})^2
\;\;\;{\rm MeV}
\ee
at a radius where $0.5\omega_{\rm bb} \la \omega_{c,e} \la \omega_{\rm bb}$.
Below this energy, the passively cooling particles emit a hard
spectrum $\omega L_\omega \sim \omega^{1/2}$.  

The {\it isotropic} power in this high-energy synchrotron
tail can be estimated.  The positrons that are created in situ
will supply the entire current --
and screen the accelerating electric field --
when their density $n_{e^+}$ satisfies
$n_{e^+}\sigma_{\rm res} r \sim \Delta\phi_{\rm N-S}$.  Here
$\sigma_{\rm res}$ is the cross-section for scattering by a
non-relativistic charge, and
$\Delta\phi_{\rm N-S}$ is the net twist angle of the field lines
that extend to $\sim 100$ km.
The corresponding power in the high-energy component is 
$L_{\rm synch} \sim 4\pi r^2 n_{e^+}(e\Delta \Phi) c$.
One sees, from eq. (\ref{potent}), that this isotropic power is 
at most comparable to the luminosity $\omega L_\omega$ at $\sim 1$ keV,
\be
{L_{\rm synch}\over (\omega L_\omega)_{\omega_{c,e}}}
 \sim \sigma_{\rm res} n_{e^+} r \la \Delta\phi_{\rm N-S}.
\ee
Thus, the output in 100 keV synchrotron photons 
is stronger in sources with stronger magnetospheric twists and
increased $\dot{P}$.

The bolometric emission of the AXP 1E 1841$-$045
is dominated by a hard continuum which rises above 20 keV 
to at least $\sim 100$ keV (Kuiper et al. 2004).   
The unpulsed spectrum diverges as $\nu F_\nu \propto \nu^{1/2}$,
and the pulsed fraction becomes close to unity at $\sim 100$ keV.  
This spectrum is suggestive of the emission of
passively cooling, relativistic particles in a magnetic field
whose cyclotron energy is less than $\sim 10$ keV.  We conclude
that the observed high energy emission is consistent with this
model only if it is somewhat more collimated than the
2-10 keV emission (within $\sim 1$ steradian).   
(The large pulsed fraction
observed at high energies is consistent with this requirement.)
This inference is testable by observing other sources:
if the output at 100 keV is consistently higher than in the 2-10
keV band, we conclude that O-mode free-free emission in a heated 
surface layer is the most plausible mechanism.  

It should be noted, in this regard, that a hard X-ray spectrum
with a $\sim -1.5$ to $-2$ photon index between $\sim 20$ and $100$ keV
has recently been detected from SGR 1806-20 by Integral (Mereghetti
et al. 2005).  However,
the $2-10$ keV spectrum of this source is harder than it is
in the case of 1E 1841$-$045, and it is possible that this is a single
power-law component that masks a yet harder bremsstrahlung component above
100 keV.  More generally, the hardness of the $2-10$
keV spectrum is observed to be correlated with overall burst
activity and with spindown rate in magnetar sources
(Marsden \& White 2001).  In particular, during
the activity of 1E 2259$+$586, the $2-10$ keV hardness was observed
to vary smoothly over the full range of values detected in the AXP
sources (Woods et al. 2004).  This pattern of spectral behavior points
to second-order Fermi acceleration of the keV black body photons
from the surface of the star, and can be effected by multiple
resonant cyclotron scattering by magnetospheric electrons (TLK).
We examine the implications of magnetospheric
beam instabilities for the formation of the $2-10$ keV spectrum
elsewhere.

In conclusion, we have described two simple mechanisms for producing
a rising energy spectrum near 100 keV in the persistent emission
of the AXP 1E 1841$-$045.  This is one of several magnetar
sources which show a hard $2-10$ keV spectrum, but otherwise little
or no evidence for X-ray burst activity.
The detection therefore provides strong evidence for a persistent 
electric current that is many orders of magnitude larger than the 
current which powers the spindown of these peculiar neutron stars.

\acknowledgements  CT is supported by the NSERC of Canada.
AMB thanks CITA for its hospitality when part of 
this work was done, and acknowledges support by the Alfred P. Sloan Foundation.

\end{document}